\documentclass[twocolumn]{aastex63}
\hypersetup{linkcolor=red,citecolor=green,filecolor=cyan,urlcolor=magenta}

\usepackage{color}
\definecolor{blue}{rgb}{0.0,0.0,1.0}
\definecolor{green}{rgb}{0.0,1.0,0.0}
\definecolor{red}{rgb}{1.0,0.0,0.0}
\definecolor{burgundy}{rgb}{0.502,0.0,0.125}
\newcommand{\benoit}[1]{{#1}} 
\newcommand{\yuhong}[1]{{#1}} 
\newcommand{\andrei}[1]{{#1}}
\newcommand{\editors}[1]{{#1}}

\received{}
\revised{}
\accepted{}
\submitjournal{ApJ}

\shorttitle{Validation of \pdfiss inversions}
\shortauthors{Afanasyev et al.}
\graphicspath{{./}{figures/}}
\newcommand\pdfiss{PDFI\_SS }

\begin{document}

\title{Validation of the \pdfiss method for electric field inversions using a magnetic flux emergence \editors{simulation}}


\correspondingauthor{Andrey N. Afanasyev}
\email{andrei.afanasev@colorado.edu}

\author{Andrey N. Afanasyev}
\altaffiliation{DKIST Ambassador}
\affiliation{Laboratory for Atmospheric and Space Physics, University of Colorado Boulder \\
1234 Innovation Drive, Boulder, CO 80303, USA}
\affiliation{National Solar Observatory, University of Colorado Boulder, Boulder, CO, USA}
\affiliation{Institute of Solar-Terrestrial Physics of SB RAS, Irkutsk, Russia}

\author{Maria D. Kazachenko}
\affiliation{Department of Astrophysical and Planetary Sciences, University of Colorado Boulder, Boulder, CO, USA}
\affiliation{National Solar Observatory, University of Colorado Boulder, Boulder, CO, USA}

\author{Yuhong Fan}
\affiliation{High Altitude Observatory, National Center for Atmospheric Research, Boulder, CO, USA}

\author{George H. Fisher}
\affiliation{Space Sciences Laboratory, University of California Berkeley, Berkeley, CA, USA}

\author{Benoit Tremblay}
\affiliation{Laboratory for Atmospheric and Space Physics, University of Colorado Boulder \\
1234 Innovation Drive, Boulder, CO 80303, USA}
\affiliation{National Solar Observatory, University of Colorado Boulder, Boulder, CO, USA}
\affiliation{Department of Astrophysical and Planetary Sciences, University of Colorado Boulder, Boulder, CO, USA}


\begin{abstract}
Knowledge of electric fields in the photosphere is required to calculate the electromagnetic energy flux through the photosphere and set up boundary conditions for data-driven magnetohydrodynamic (MHD) simulations of solar eruptions. Recently, the \pdfiss method for inversions of electric fields from a sequence of vector magnetograms and Doppler velocity measurements was improved to incorporate spherical geometry and a staggered-grid description of variables. The method was previously validated using synthetic data from anelastic MHD (ANMHD) simulations. In this paper, we further validate the \pdfiss method, using approximately one--hour long MHD simulation data of magnetic flux emergence from the upper convection zone into the solar atmosphere. We reconstruct photospheric electric fields and calculate the Poynting flux, and compare those to the actual values from the simulations. We find that the accuracy of the \pdfiss reconstruction is quite good during the emergence phase of the simulated ephemeral active region evolution and decreases during the shearing phase. Analysing our results, we conclude that the more complex nature of the evolution (compared to the previously studied ANMHD case) that includes the shearing evolution phase is responsible for the obtained accuracy decrease.
\end{abstract}

\keywords{Solar magnetic flux emergence  --- Solar active region magnetic fields --- Solar active region velocity fields  --- Solar active regions --- Solar photosphere --- Solar atmosphere}

\section{Introduction} \label{sec:intro}

Although it is the magnetic field and magnetic energy stored in various topological structures that are believed to cause all the phenomena of solar \editors{transient} activity, the information on both magnetic and electric fields is required to estimate electromagnetic energy fluxes through the solar atmosphere.
Those fluxes provide valuable information on the energy transported from the solar interior to the atmosphere, contributing to the understanding of magnetic heating of the higher solar atmosphere (e.g. \citealt{Welsh-pasj-2015}) as well as to the understanding of physical processes in solar eruptions and flares. In particular, electric fields at the photosphere are used to set up boundary conditions for data-driven magnetohydrodynamic (MHD) simulations of the emergence of a new magnetic flux, formation of flux ropes, and subsequent eruptions (e.g. \citealp{2004ApJ...609.1123F, Fan_2010, Toriumi_2020, Hoeksema_2020ApJS..250...28H}, see \citealp{Cheung_2014LRSP...11....3C} for a review).

Because direct observations of electric fields based on the Stark effect are quite difficult to achieve due to the low sensitivity of measurements (e.g. \citealp{Foukal_1995SoPh..156..293F}), much effort was put into development of the techniques to estimate photospheric electric fields using other measured quantities. Several methods (\citealp{Kusano-2002ApJ...577..501K, FLCT2008ASPC..383..373F, Schuck_2008ApJ...683.1134S}, see \citealp{Welsch_2007} for comparison) estimated the plasma velocities in the photosphere to obtain the electric field from the ideal Ohm's law ($\mathbf{E} \propto \mathbf{V} \times \mathbf{B}$). These methods generally track changes in the magnetic field from a sequence of magnetograms, although some of them also used the vertical component of Faraday's law ($\partial \mathbf{B} / \partial t \propto \nabla \times \mathbf{E}$) as an additional constraint. The weakness of such approaches lay in extracting the information about apparent (or optical) flows, which could be affected by the temporal evolution of magnetic fluxes, e.g. their emergence or submergence, and therefore did not correspond to real plasma flows in the photosphere. That affected the quality of the reconstructed electric field, leaving the problem unresolved. The recent development of techniques using neural networks, however, promises a significant improvement in plasma velocity estimating methods, which could provide accurate estimates of the photospheric electric fields.  In particular, neural networks such as DeepVel \citep{Asensio2017, Tremblay-2018SoPh..293...57T} and DeepVelU \citep{Tremblay-2020FrASS...7...25T} reconstruct plasma velocities from surface observations by emulating flows in realistic MHD simulations of the photosphere. Note that flows estimated using DeepVel and DeepVelU are neither optical nor physical as the neural networks currently do not solve an optical flow equation or physical equations.

A different approach to estimate electric fields was proposed by \citealt{Fisher-2010-ApJ...715..242F}. The idea was to uncurl Faraday's law, using the poloidal-toroidal decomposition (PTD) for the magnetic field time derivative. The solution of the obtained set of equations for the electric field was not unique, however, and the gradient of any scalar function could be added into the inductive electric field, with the resulting field still satisfying the induction condition. In particular, the pure inductive electric field was not perpendicular to the magnetic field, as required by the ideal Ohm's law. That fact motivated \citealt{Fisher-2012-SoPh..277..153F} to include the additional observational information about the line-of-sight (LOS) component of the plasma velocity from photospheric Dopplergrams as well as about horizontal optical flows obtained using the Fourier Local Correlation Tracking (FLCT) technique \citep{FLCT2008ASPC..383..373F} \benoit{on a sequence of LOS magnetograms}. Those data provided contributions into the potential component of the electric field. \citet{Yeates_2017} and \citet{Pomoell-2019SoPh..294...41P} also pointed out the importance of the non-inductive electric fields. The perpendicularity between the resulting electric field and magnetic field in the \pdfiss method was finally achieved with an iterative technique eliminating the electric field component parallel to the magnetic field vector. On the other hand, the iterative technique appeared to provide the correct use of \benoit{optical} horizontal flows as an available proxy, instead of real horizontal plasma velocities. In addition, the iterative technique allowed other missing observational information to be taken into account phenomenologically.

The new method was later referred to as the PDFI method for the techniques and data used for electric field estimates: PTD -- Doppler -- FLCT -- Ideal, where ``Ideal'' stood for the iterative technique to get perpendicularity of the electric and magnetic fields, coming from Ohm's law for ideal plasmas. \citet{Masha-2014ApJ...795...17K} significantly expanded the scope of the PDFI method by including non-normal viewing angles for observations, improved performance with the use of the FISHPACK library, and performed a test to validate the method. More specifically, the results of the electric field inversions based on synthetic magnetograms from anelastic magnetohydrodynamic (ANMHD) simulations \citep{ANMHD_2004JASTP..66.1257A} were compared to the actual ANMHD electric fields. \citet{Masha-2014ApJ...795...17K} found the best values for free parameters in the PDFI method and applied it to real solar data to study the evolution of the energy and magnetic helicity fluxes in a solar active region \citep{Masha-2015ApJ...811...16K}. \citet{Fisher2020-ApJS..248....2F} further improved the PDFI method by including spherical geometry as well as switching to a more conservative staggered-grid description of variables, hence the new method name became \pdfiss. 

So far, the ANMHD simulation data were the only synthetic test data used for validation of the \pdfiss software.
However, as mentioned in \citet{Masha-2014ApJ...795...17K}, the vertical emergence of the bipole magnetic flux was probably exaggerated in the synthetic data from the ANMHD simulations, because the simulations are limited to modelling the rise of the flux tube in the convection zone without modelling the emergence process into the solar atmosphere. Both observations \citep[e.g.][]{Liu-Schuck2012ApJ...761..105L} and MHD simulations of flux emergence into the solar atmosphere \citep[e.g.][]{Manchester:etal:2004, Fan:2009} have shown that photospheric shearing motions in the emerging region contribute significantly to the free-energy and helicity accumulation in the corona, which are largely absent in the synthetic test data from the ANMHD simulations.
In addition, the ANMHD data were smooth in time and space, thus not revealing small-scale structures or small-scale flow patterns that could potentially affect the reconstructed electric field in a significant way. Recently, \citet{Lumme-2019SoPh..294...84L} performed an analysis of the dependence of the \pdfiss reconstructed energy and magnetic helicity fluxes on the temporal cadence of the input data, using observations of a solar active region, and found that the \pdfiss results were stable for a range of the sampled dataset cadences from several minutes up to two hours.

In this paper, we test the \pdfiss electric field inversion method, using synthetic photospheric data obtained from an MHD simulation \yuhong{of a twisted magnetic flux tube emerging from the top layer of the convection zone into the atmosphere and the corona}, similar to the simulation in \citet{Fan:2009}, where photospheric shearing and rotational motions of the polarities develop to drive the build-up of a corona flux rope. We apply the \pdfiss method to reconstruct the photospheric electric field from a sequence of the synthetic photospheric magnetograms and vertical velocity data extracted from the simulation, and compare the results to the actual electric field in the MHD simulation.

The paper is organised as follows. In Section~\ref{sec:simulationdata}, we first give a model description of the flux emergence simulation, and then analyse the synthetic photospheric data extracted from the simulation in Section~\ref{sec:mfe}. In Section~\ref{sec:flct}, we perform FLCT reconstructions of the horizontal optical flows from simulated magnetograms of the vertical component of the photospheric magnetic field.  In Section~\ref{sec:pdfiss}, we show \pdfiss reconstructions of the photospheric electric field and discuss our results. Section~\ref{sec:conclusion} presents our conclusions.

\section{Magnetic flux emergence simulation} \label{sec:simulationdata}

Here we give a description of the \yuhong{interior-to-corona simulation of the emergence of a twisted magnetic flux tube}, of which the photospheric magnetic and velocity field data are used in this work for the validation of the \pdfiss inversion method. The detailed results of the flux emergence simulation will be presented in a separate paper.

The simulation uses the ``Magnetic Flux Eruption'' (MFE) code, which solves the following semi-relativistic magnetohydrodynamic equations \citep{Fan:2017}:
\begin{equation}
\frac{\partial \rho}{\partial t}
= - \nabla \cdot ( \rho {\bf v}) ,
\label{eq:cont}
\end{equation}
\vspace{-7mm}
\begin{eqnarray}
\frac{\partial ( {\rho \bf v})}{\partial t}
&=& - \nabla \cdot \left ( \rho {\bf v} {\bf v} \right )
- \nabla p + \rho {\bf g} + \frac{1}{4 \pi}
( \nabla \times {\bf B} ) \times {\bf B}
\nonumber \\
& & + \frac{{v_A}^2/c^2}{1 + {v_A}^2/c^2} \left[ {\cal I}
- {\hat{\bf b}} {\hat{\bf b}} \right ] \\
& & \cdot \left [ ( \rho {\bf v} \cdot \nabla ) {\bf v} + \nabla p  - \rho {\bf g} - \frac{1}{4 \pi} ( \nabla \times {\bf B} ) \times {\bf B} \right ], \nonumber
\label{eq:motion}
\end{eqnarray}
\begin{equation}
\frac{\partial {\bf B}}{\partial t}
= \nabla \times ({\bf v} \times {\bf B}),
\label{eq:induc}
\end{equation}
\begin{equation}
\nabla \cdot {\bf B} = 0,
\label{eq:divb}
\end{equation}
\begin{equation}
\frac{\partial e}{\partial t} = - \nabla \cdot
\left ( {\bf v} e \right ) - p \nabla \cdot {\bf v}
- \nabla \cdot {\bf q}_s + Q_{\rm rad} + H_{\rm em} + H_{\rm num}  ,
\label{eq:energy}
\end{equation}
\begin{equation}
p = \frac{\rho R T}{\mu},
\label{eq:state}
\end{equation}
where
\begin{equation}
e = {\frac{p}{\gamma - 1} },
\label{eq:eint}
\end{equation}
\begin{equation}
v_A = \frac{B}{\sqrt {4 \pi \rho}} .
\label{eq:alfvenspeed}
\end{equation}
Here the equations are solved in Cartesian geometry. In the above equations, symbols have their usual meanings, where ${\bf v}$ is the velocity field, ${\bf B}$ is the magnetic field and $B$ is its magnitude, $\rho$, $p$, and $T$ are respectively the plasma density, pressure and temperature, $e$ is the internal energy density, $c$ is the (reduced) speed of light, ${\cal I}$ is the unit tensor, ${\hat {\bf  b}} = {\bf B} / B$ is the unit vector in the magnetic field direction, ${\bf g} = -g {\hat {\bf z}} $ is the constant gravitational acceleration, $R$, $\mu$, and $\gamma$, are respectively the gas constant, the mean molecular weight, and the adiabatic index of the perfect gas. We assume the adiabatic index ${\gamma} = 5/3$. The internal energy equation (Eq.~\ref{eq:energy}) takes into account the following non-adiabatic effects:
i) the field-aligned heat conduction flux, ${\bf q}_s$, which we compute using the hyperbolic heat conduction approach \citep{Rempel:2017};
ii) the optically thin radiative cooling, $Q_{\rm rad}$:
\begin{equation}
Q_{\rm rad} = N^2 \Lambda (T) ,
\label{eq:radloss}
\end{equation}
where $N = \rho / m_p$ is the proton number density assuming a fully ionised hydrogen gas, with $m_p$ being the proton mass, and the radiative loss function $\Lambda (T)$ is as given in \citet{Athay:1986}, with the cooling suppressed
for $T$ below $2 \times 10^4$ K;
iii) an empirical coronal heating, $H_{\rm em}$, that is a function of height $z$:
\begin{equation}
H_{\rm em} = \frac{F}{L_H} \exp \left[ -(z - z_0) / L_H \right ] ,
\label{eq:heatingfunc}
\end{equation}
which is non-zero for $z>z_0$,
with $F=5 \times 10^5 \; {\rm erg \; s^{-1} cm^{-2}}$, $L_H= 11 \; {\rm Mm}$,
and $z_0 = 3 \; {\rm Mm}$ above the photosphere.
Finally, we also evaluate the effective heating resulting from numerical diffusion of velocity and magnetic fields in the momentum and induction equations, and added it to the internal energy in Equation~(\ref{eq:energy}) as $H_{\rm num}$.

\begin{figure}[htb!]
\centering
\includegraphics[width=0.47\textwidth]{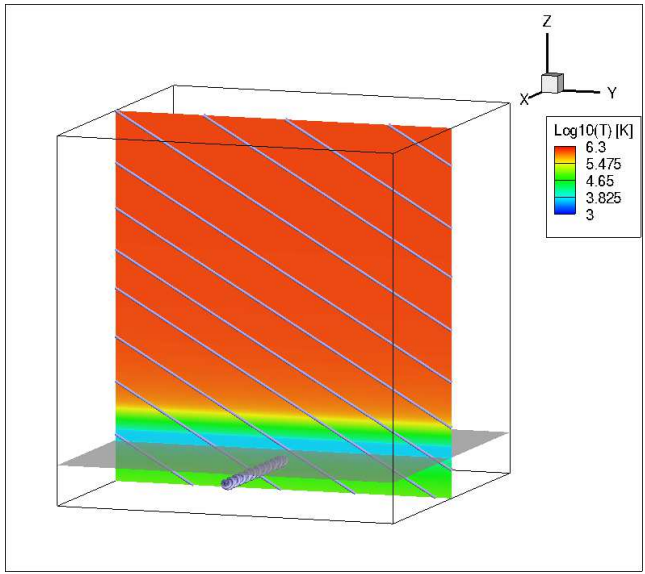}
\caption{The initial setup of the simulation domain. The grey plane is the photosphere. The colour image in the vertical cross-section shows the temperature of the initial background equilibrium plane-parallel atmosphere permeated with a uniform open magnetic field represented by the field lines in the cross-section. A twisted magnetic flux tube (as represented by the twisted field lines) \editors{oriented along the $x$-axis} is inserted into the interior layer below the photosphere. See text for detailed description.}
\label{fig:domain}
\end{figure}

\begin{figure}
\centering
\includegraphics[width=0.49\textwidth]{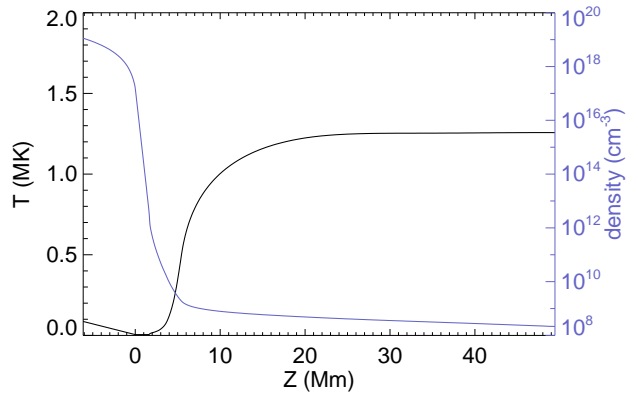}
\caption{The temperature (black curve) and density (blue curve) profiles of the initial equilibrium plane-parallel atmosphere.}
\label{fig:initstr}
\end{figure}

The setup of the simulation is as follows. We use a Cartesian domain as shown in Figure~\ref{fig:domain} with the horizontal sizes:
$x \in [-26.5, 26.5] \;{\rm Mm}$,
$y \in [-26.5, 26.5] \;{\rm Mm}$,
and the vertical size:
$z \in [-6.1, 49.8] \;{\rm Mm}$, where $z=0$ corresponds to the photosphere (the grey horizontal plane in Figure~\ref{fig:domain}).
The domain is resolved with a non-uniform grid of $480(x) \times 480(y) \times 512(z)$ cells, with a peak vertical resolution of $\Delta z = 31$~km at the photosphere and a peak horizontal resolution of $\Delta x = \Delta y = 46$~km centred on the emerging region.
We first initialise a plane-parallel, vertically stratified atmosphere, as shown in Figure~\ref{fig:initstr}, which \yuhong{consists of an adiabatically stratified \andrei{polytropic layer} representing the top of the convection zone, an approximately isothermal layer that represents the photosphere and chromosphere, \andrei{and} a high temperature isothermal layer representing the corona}, permeated with a uniform slanted open magnetic field of 6~G (illustrated by the field lines shown in the vertical cross-section in Figure~\ref{fig:domain}).
This initial one-dimensional atmosphere with a uniform open magnetic field has been numerically relaxed to an approximate hydrostatic and thermal equilibrium along the magnetic field, with the transition region and hot corona maintained by the empirical coronal heating (Eq.~\ref{eq:heatingfunc}), balanced by the optically thin radiative cooling and the field-aligned heat conduction. A horizontal twisted magnetic flux tube is inserted into the convection zone layer below the photosphere (see Figure~\ref{fig:domain}), whose magnetic field is given as follows:
\begin{equation}
{\bf B} = B_x (r) \, \hat{\bf x} + B_{\theta} (r) \, \hat {\bf \theta},
\end{equation}
where
\begin{equation}
B_x (r) = B_{0} \exp (- r^2 /a^2 ) ,
\end{equation}
\begin{equation}
B_{\theta} (r) = q r B_x (r) ,
\end{equation}
$\hat{\bf x}$ is the tube axial direction, $\hat{\bf {\theta}}$ denotes the azimuthal direction in the tube cross-section, $r$ denotes the radial distance to the tube axis which is located at $(y,z) = (0, -1.69) \;{\rm Mm}$, $a=0.38$~Mm is the radius of the tube, $q$ is the rate of the field-line twist and is set to be $-a^{-1}$, and $B_0 = 5677 $ G is the field strength at the initial axis of the flux tube. The total flux of the subsurface flux tube is $2.63 \times 10^{19}$~Mx, much smaller than the flux of typical active regions, but corresponds to the flux scale of ephemeral regions \citep[e.g.][]{Hagenaar:etal:2008}. Inside the flux tube, the gas pressure differs from that of the initial background plane parallel atmosphere $p_0 (z)$ by:
\begin{equation}
p_1 (r) = - \frac{B_0^2}{2} \exp (- 2 r^2 / a^2 )
\left [ 1 - \frac{1}{2} q^2 a^2 \left ( 1 - \frac{2r^2}{a^2}
\right ) \right ],
\label{p1}
\end{equation}
such that the pressure gradient from $p_1$ balances the Lorentz force of the twisted flux tube. Furthermore, the middle segment of the flux tube is made buoyant by introducing a density change $\rho_1$ within the flux tube relative to the density $\rho_0(z)$ of the background plane parallel atmosphere:
\begin{equation}
\rho_1 = - \rho_0(z) \frac{B_x^2 (r)}{2 p_0(z)}
\left [ (1+\epsilon) \exp ( - x^2 / \lambda^2 ) - \epsilon \right ]
\end{equation}
where the length scale of the buoyant segment $\lambda = 1.54$ Mm,
and $\epsilon = 0.1$.

\begin{figure}[htb!]
\centering
\includegraphics[width=0.47\textwidth]{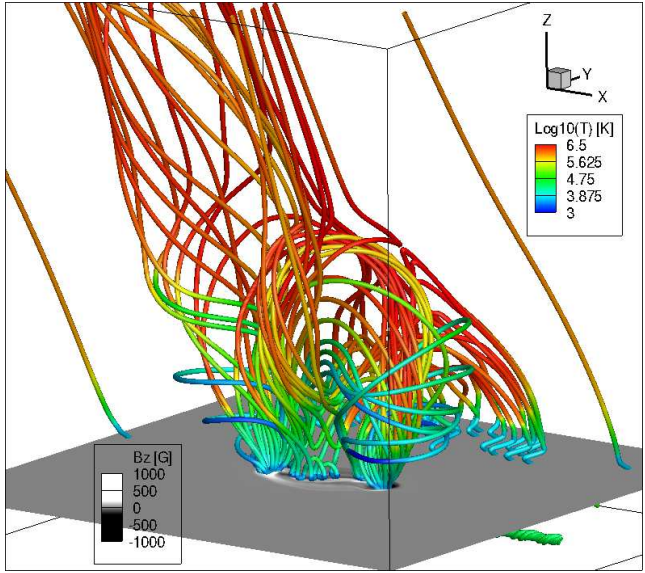}
\caption{3D magnetic field at $t=45$~min from the flux emergence simulation, \benoit{more specifically during the shearing phase}. The field lines are coloured by temperature.  The horizontal cross-section corresponds to the photosphere with grey-scale image showing the vertical magnetic field.}
\label{fig:coronalfluxrope}
\end{figure}

Given the initial state described above, the subsurface flux tube develops an $\Omega$-shape, with its central buoyant segment rising towards the photosphere. When the apex of the flux tube reaches the stably stratified photosphere, \yuhong{the magnetic buoyancy instability sets in, \editors{which causes the upper boundary of the flux tube to expand into the atmosphere and begins the flux emergence process}}, similar to what has been found in many previous simulations \citep[see e.g.][]{Manchester:etal:2004, Fan:2009, Archontis:Hood:2013}. In this case, as a result of the flux emergence, a twisted coronal flux rope is built up in the corona as shown in Figure~\ref{fig:coronalfluxrope}.
The flux rope reconnects with the ambient open field, leading to the development of a rotating ``blow-out'' jet with untwisting field lines in the jet column (see Figure~\ref{fig:coronalfluxrope}). The detailed result of the coronal magnetic field evolution resulting from the flux emergence will be described in a separate paper (Manek et al. in preparation). It is found that the formation of the coronal flux rope is not due to the bodily emergence of the subsurface flux rope, but is mainly due to the shearing and rotational motions of the polarity concentrations at the photosphere during the flux emergence, which transport twist from the subsurface flux tube into the corona \citep[e.g.][]{Fan:2009}.  In the following section, we extract the time sequence of the photospheric vector magnetic field and velocity field evolution from the flux emergence simulation as synthetic data, and apply the \pdfiss method to the synthetic magnetograms to examine how well it infers the electric field at the photosphere.

\begin{figure*}
\centering
\includegraphics[width=0.99\hsize]{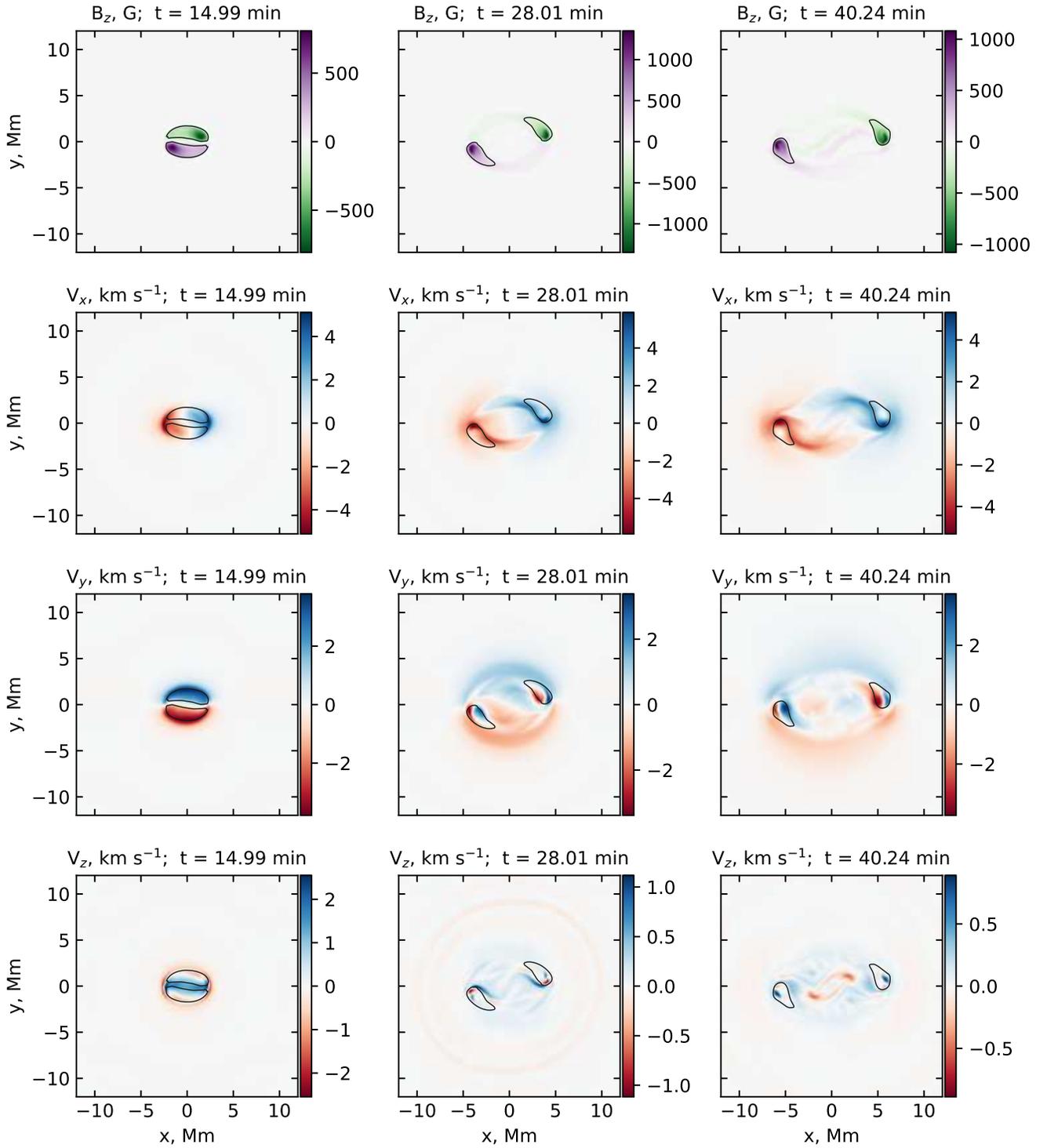}
\caption{Slices of the solar surface in the MFE simulation representing the  vertical component of the magnetic field and the three components of the plasma velocity at three different times: \andrei{14.99~min} (left column), \andrei{28.01~min} (middle), and \andrei{40.24~min} (right). The black contours are drawn for values of $\pm 200$~G of the vertical component of the magnetic field. 
\label{fig_mag_n_velocity_fields}}
\end{figure*}
%
\begin{figure*}[tb]
\centering
\includegraphics[width=0.49\hsize]{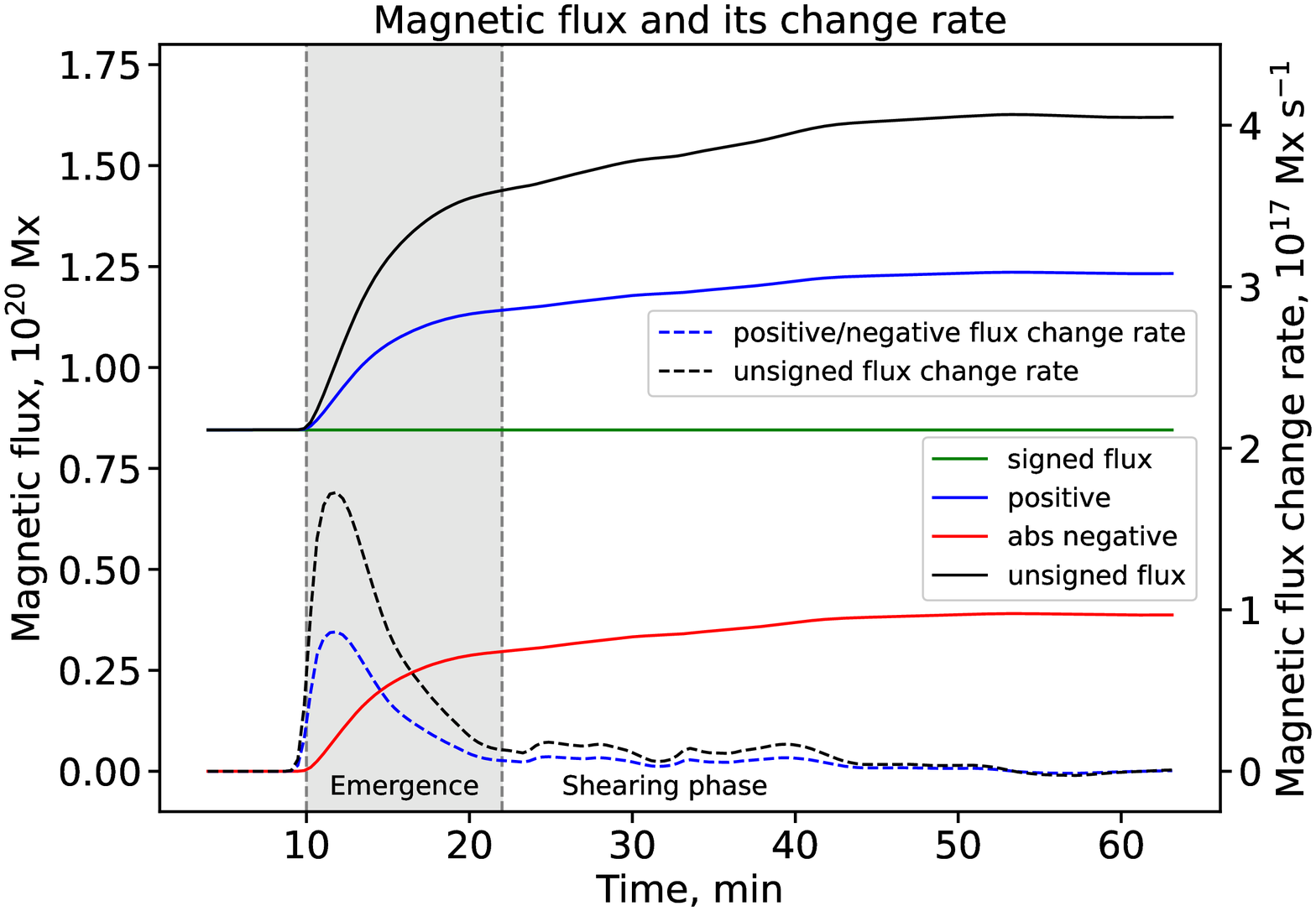}
\includegraphics[width=0.49\hsize]{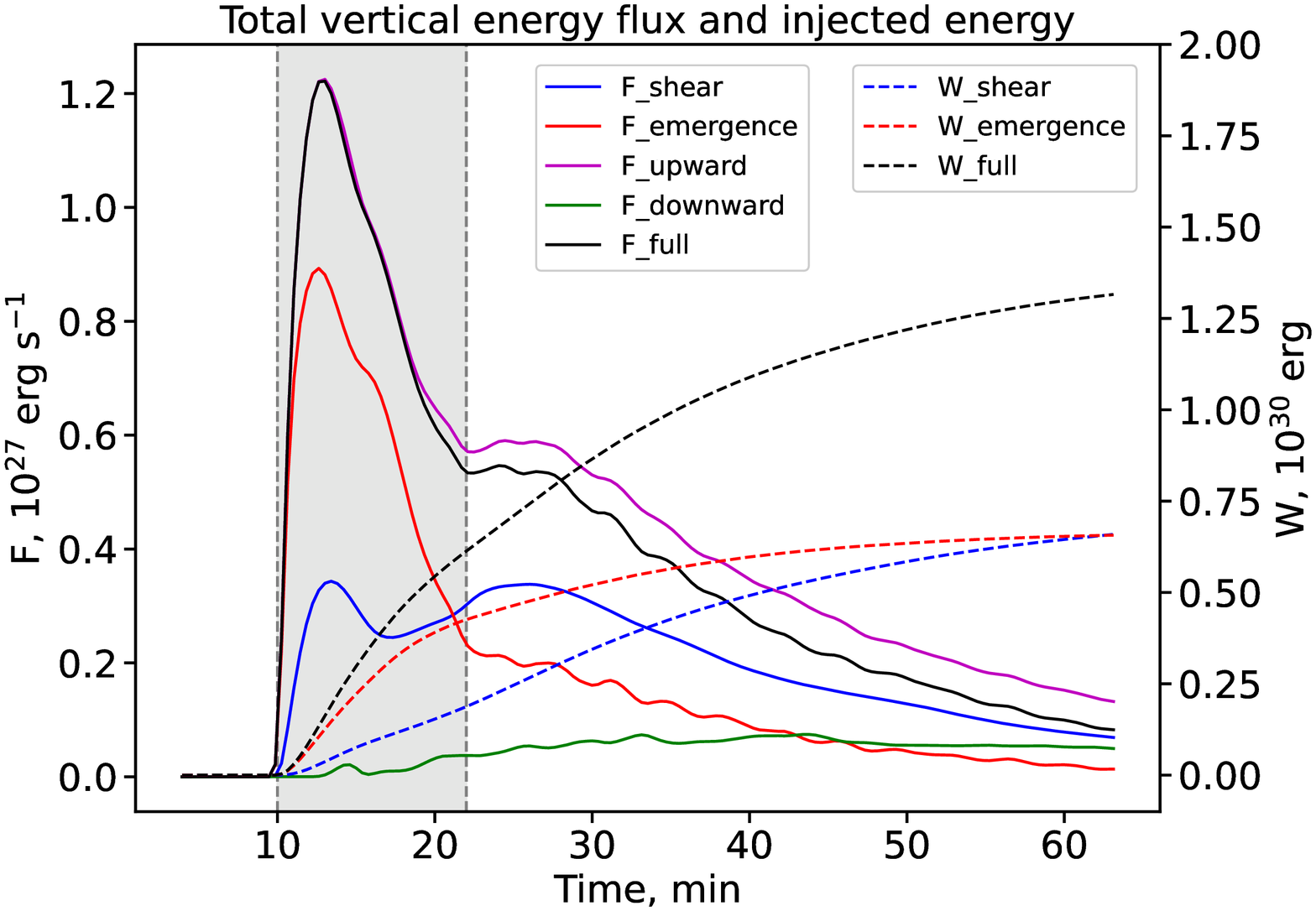}
\caption{Analysis of the MFE simulated emergence of a new magnetic flux into the solar atmosphere. The left panel shows the evolution of the magnetic flux through the computational domain and its change rate. The solid lines represent the unsigned magnetic flux (black), signed (green), positive (blue), and absolute value of the negative magnetic flux (red). The dashed lines represent the change rates for the unsigned (black), and positive and negative (blue) magnetic fluxes. The right panel shows the evolution of the total energy flux $F$ through the computational domain and the injected energy $W$. The solid lines represent the shear component (blue), emergence component (red), and full energy flux (black). The purple and green solid lines demonstrate the upward and downward contributions to the full energy flux, respectively. The dashed lines represent the amount of the injected energy as calculated from the full energy flux (black), and the emergence (red) and shear (blue) components of the energy flux. The axes for dashed lines are located on the right side of each panel. The shaded areas in both panels represent the emergence phase (10--22~min) of the simulated active region's evolution. \benoit{The second dashed vertical line of each plot marks the beginning of the shearing phase ($t > 22$ min).}
\label{fig_mflux_and_energy}}
\end{figure*}
%
\begin{figure*}
\centering
\includegraphics[width=0.98\hsize, trim={0cm 0 0 0}, clip]{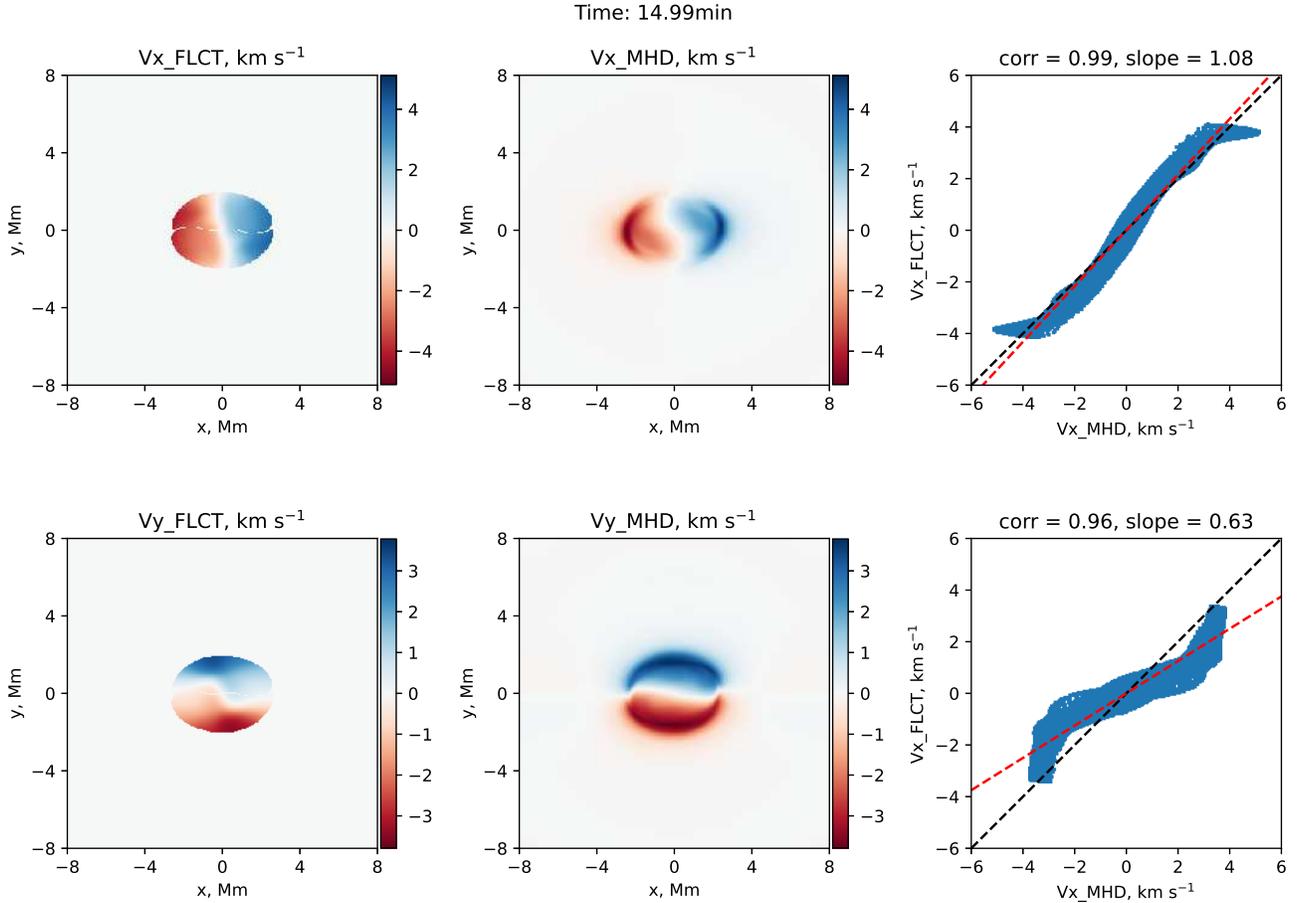}
\caption{Comparison between the horizontal $x$- and $y$-components (top and bottom rows, respectively) of the plasma velocity reconstructed using FLCT (left column) and computed by the MFE simulation (middle column) at \benoit{$t=14.99$~min}, \benoit{\textit{i.e.} during the emergence phase of the simulation}. The right panels show the scatter plots, linear fit (red dashed line), Spearman correlation coefficient and slope of the fit, and reference one-to-one line (black dashed).
\label{fig_flct_snapshots_t15}}
\end{figure*}
%
\begin{figure*}
\centering
\includegraphics[width=0.98\hsize, trim={0cm 0 0 0}, clip]{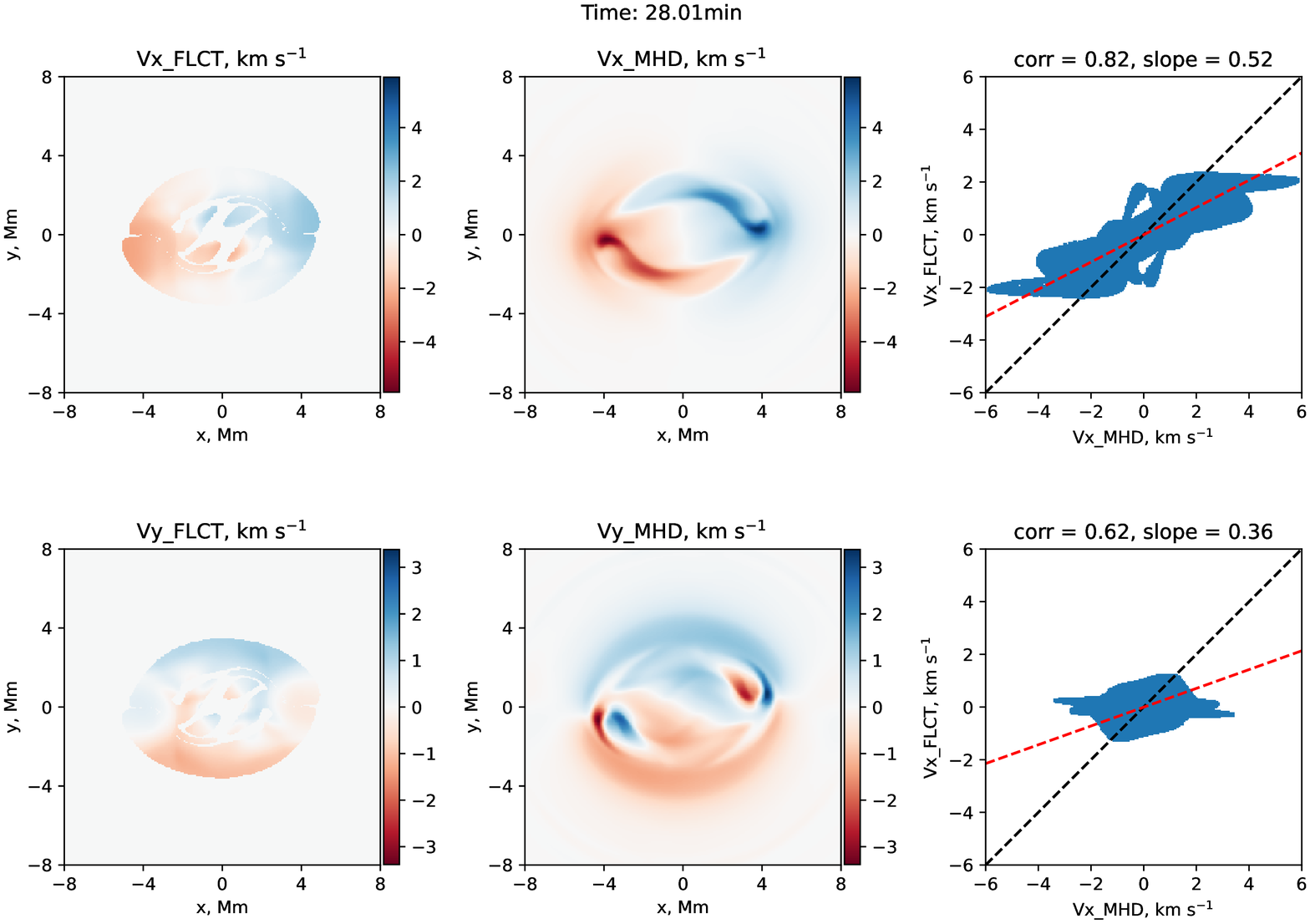}
\caption{The same as Figure~\ref{fig_flct_snapshots_t15} at \benoit{$t=28.01$~min}, \benoit{\textit{i.e.} during the shearing phase of the simulation}.
\label{fig_flct_snapshots_t28}}
\end{figure*}

\section{Analysis of the photospheric synthetic data} \label{sec:mfe}

We use simulation data for the magnetic and velocity fields at the photospheric level and the ideal Ohm's law to calculate our reference electric fields for the validation of \pdfiss. The dataset spans approximately one hour in time. After a 10--min initial stage of the flux emergence, the two polarities of the bipolar region start moving in the opposite directions. Figure~\ref{fig_mag_n_velocity_fields} shows the typical patterns for the vertical component of the photospheric magnetic field as well as for the components of the plasma velocity at three different times. The times have been selected to demonstrate different stages of the emerging magnetic flux evolution. An important feature of the simulated data is the shear structures seen in the 28 and 40~min snapshots for \andrei{both horizontal components} of the plasma velocity in the regions corresponding to the magnetic polarities. \andrei{The vertical} magnetic field does not reveal \andrei{such \editors{shearing} structures, although showing some fine swirling dynamics.} These shear structures appear to be associated with rotational motions of the twisted emerging flux, which are detected in solar white-light images as rotating sunspots (e.g. \citealp{Masha-2009ApJ...704.1146K}). We also note that vertical (Doppler) velocities decrease with time. \andrei{The late stage of the simulated evolution contains downflow regions in the middle of the active region, which are caused by the draining and tilting motions of U-shaped, anchored field lines of the partially emerged flux rope.}

Figure~\ref{fig_mflux_and_energy} analyses the evolution of the flux and energy characteristics of the MFE simulated active region. We calculate the magnetic flux and its change rate through the photospheric cross-section of the MFE computational domain. \andrei{The photospheric magnetic flux is a combination of the emerging sub-surface flux tube and a uniform slanted open magnetic field permeating the simulation domain.} \yuhong{The open field flux is of positive polarity on the photosphere. We note that the (absolute) value of the negative polarity flux on the photosphere (due to flux emergence) increases from zero to a peak value of about $3.9 \times 10^{19} $Mx (red solid curve in left panel of Figure~\ref{fig_mflux_and_energy}), significantly exceeding the total axial flux of the subsurface flux tube ($2.63 \times 10 ^{19}$ Mx). This is because some of the twisted field lines of the partially emerged flux tube thread through the photosphere with more than one stitch, and thus the unsigned emerged flux of each polarity on the photosphere can exceed the total axial flux of the subsurface twisted flux tube.}

\andrei{In the left panel of Figure~\ref{fig_mflux_and_energy},} we can distinguish two phases of the magnetic flux evolution. The first one (around 10--22~min) is characterised by a rapid change in the magnetic flux due to the flux emergence, which is also observed in the magnetic flux change rate. After 22~min, the magnetic flux does not increase so significantly. Throughout the simulation, the signed flux stays constant because of the symmetry in the flux variations at both polarities, with its non-zero value being due to the constant background magnetic field of 6~G.

Analysing the energy fluxes, we break the vertical component of the Poynting flux, $S_z$, into two \editors{terms}, 
\begin{equation}
S_z = \frac{c}{4 \pi} \left[\mathbf{E} \times \mathbf{B} \right]_z = 
\frac{B_h^2}{4 \pi} v_z - \frac{B_z}{4 \pi} \left( \mathbf{B}_h \cdot \mathbf{v}_h \right), 
\label{eq-pointing-flux}
\end{equation}
where $\mathbf{E}$ is the electric field vector, and the subscripts stand for the horizontal (h) and vertical (z) vector components \citep[e.g.][]{Abbet_Fisher_2012SoPh..277....3A}. The emergence \editors{term} determined by the vertical component of the plasma velocity corresponds to the first term in Equation~(\ref{eq-pointing-flux}), while the shearing \editors{term} is determined by the horizontal flows and corresponds to the second term in the equation. The right panel in Figure~\ref{fig_mflux_and_energy} shows that the first, rapid phase in the magnetic flux evolution corresponds to the deposit of energy into the solar atmosphere \andrei{predominantly} due to the emergence component of the energy flux \andrei{(the red solid line)}, while the quiet phase corresponds to the shearing evolution of the active region from the viewpoint of the deposited energy flux \andrei{(the blue solid line). Therefore, throughout the paper, we will refer to the first phase of the simulated active region evolution (10--22~min) as the emergence phase, while the subsequent evolution of the active region (after 22~min) will be referred to as the shearing phase (see Figure~\ref{fig_mflux_and_energy}). The shearing phase starts once the contribution to the vertical electromagnetic energy flux from the horizontal (shearing) motions dominates that from the vertical plasma motions. Physically, this is due to the decrease of vertical plasma flows during the course of simulated active region evolution \editors{because of the completion of the flux emergence process}.} The analysis of the upward and downward contributions into the full vertical energy flux demonstrates the dominant energy inflow to the atmosphere from below. We also note that cumulative contributions to the injected energy from the emergence and shearing energy flux components in the simulation are equal to each other.

\begin{figure}
\centering
\includegraphics[width=0.8\hsize, trim={0cm 0 0 0}, clip]{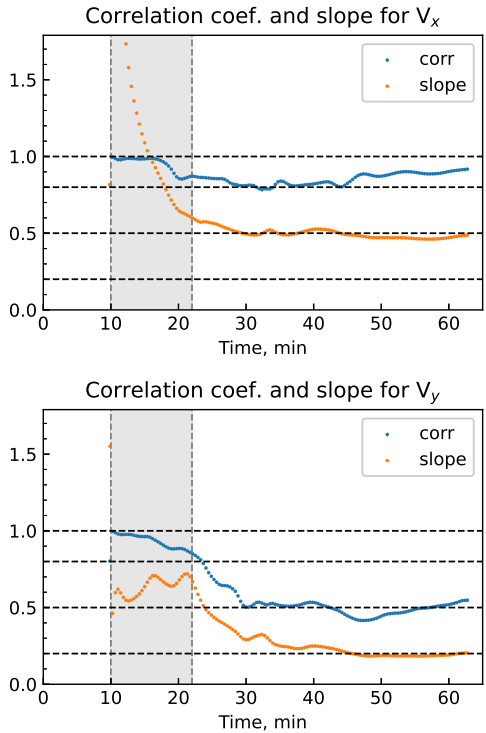}
\caption{Evolution of the Spearman correlation coefficient (blue) and slope parameter (yellow) of the horizontal velocity FLCT reconstructions for the entire duration of the simulation. Upper (lower) panel shows results for the $x$($y$)-component of the flows. The shaded areas represent the emergence phase (10--22~min) of the simulated active region evolution. \benoit{The second dashed vertical line of each plot marks the beginning of the shearing phase ($t > 22$ min)} \andrei{The horizontal dashed lines mark the limits for the strong, moderate, and weak correlation.}
\label{fig_flct_corr_slope}}
\end{figure}

\section{FLCT inversions of horizontal flows} \label{sec:flct}
To test the \pdfiss software, we first apply the FLCT technique \citep{FLCT2008ASPC..383..373F} to the MFE $B_z$~magnetograms to obtain estimates of horizontal plasma flows in the photosphere. For each timestep, we use a pair of magnetograms that are one timestep behind and ahead of the given time.

For the FLCT technique, we have to specify free parameters. To find the best value for the windowing parameter $\sigma$, we performed several FLCT reconstructions of horizontal velocities with $\sigma =$ 5, 10, 15, 20, 25, 50~pixels, and selected the optimal value $\sigma = 20$~pixels by computing the correlation between the reconstructed and reference MFE velocities (cf. $\sigma = 5$ in \citealp{Fisher-2012-SoPh..277..153F}, $\sigma = 15$ in \citealp{Masha-2014ApJ...795...17K}). We chose the threshold value of 14~G for the magnitude of the minimum magnetic field, which is about 1\% of the maximum value of $B_z$ during the entire simulation. A threshold value is often set up when working with observational data to eliminate noisy data in the weak field regions. Here we introduce the threshold to reduce the computational time and also for consistency with the previous studies. Moreover, magnetic field masks are used by the \pdfiss software when computing the Doppler contribution to the electric field \citep{Fisher2020-ApJS..248....2F}. We performed several FLCT reconstructions with other threshold values (even one order of magnitude larger) and have not found a significant effect on the results.

Figure~\ref{fig_flct_snapshots_t15} and Figure~\ref{fig_flct_snapshots_t28} compare the FLCT reconstructions of horizontal optical flows to the MFE simulation flows at two different times. We use the Spearman correlation coefficient, $c_S$, and the slope angle of the linear fit as statistical metrics, both determined from the scatter plots, and find that the quality of inversions is worse for the $y$-component of the plasma velocity \benoit{than for the $x$-component for both times and that quality of both components is worse at $t=28.01$ min than at $t=14.99$ min}. These discrepancies are most likely due to the more complex evolution of the magnetic flux emergence, in particular, due to the fact that the shear (rotational) structures are not detectable in the simulated magnetograms (see Section~\ref{sec:mfe} \benoit{and Figure \ref{fig_mag_n_velocity_fields}}) \benoit{and thus cannot be tracked by the FLCT algorithm in \andrei{both} directions}. 
\andrei{Indeed, the FLCT technique tracks changes in the structure of the $B_z$-component and provides information on the optical horizontal flows. In the case of study, we have two kinds of motions in the photospheric level: i) \editors{translational} motions of two regions of the opposite magnetic polarities, and ii) their rotational motions. However, unlike the $y$-component of the horizontal flow, its $x$-component has a significant contribution from the \editors{translational} motions of the polarities, which is captured by the FLCT technique. That is why the reconstruction of the $x$-component of horizontal photospheric flows is better during the entire course of the simulation.}

In Figure~\ref{fig_flct_corr_slope}, we show the evolution of the correlation coefficient between the reconstructed and simulation velocities and slope parameter over the total duration of the simulation. Throughout the paper, we use the following correlation scores: strong correlation ($c_S>0.8$), moderate ($0.5<c_S<0.8$), and weak ($0.2<c_S<0.5$). The FLCT reconstruction of the $y$-component of horizontal flows shows strong correlation with MFE velocities during the emergence phase of the simulated magnetic flux evolution (see Section~\ref{sec:mfe}), \benoit{and a moderate-to-weak correlation during the shearing phase which is consistent with the results from Figure~\ref{fig_flct_snapshots_t28} taken during this phase.}  
\benoit{The metrics for the reconstruction of the $x$-component of horizontal flows decrease when transitioning from the emergence phase to the shearing phase, but per our criteria the correlation remains strong throughout the entire simulation duration.} 
We should also keep in mind that the FLCT technique tracks apparent flows rather than physical plasma velocities, which could justify the lower values obtained for the slope parameter.

\begin{figure}
\centering
\includegraphics[width=0.89\hsize, trim={0cm 0 0 0}, clip]{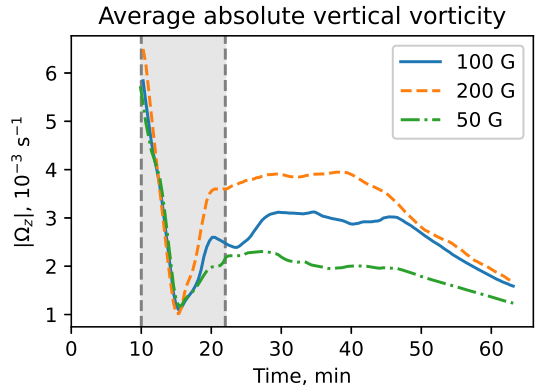}
\caption{Evolution of the average absolute vertical vorticity calculated using the MFE simulation horizontal velocities \editors{for different magnetic field mask values: 100~G (blue solid line), 200~G (yellow dashed), 50~G (green dash-dotted)}. The shaded area represents the emergence phase (10--22~min) of the simulated active region evolution. The second dashed vertical line marks the beginning of the shearing phase ($t > 22$ min).
\label{fig_vorticity}}
\end{figure}

\andrei{In order to understand better the influence of rotational plasma motions, we plot \editors{the temporal evolution of the absolute vertical vorticity, namely} the $z$-component of the curl of the photospheric flows, averaged throughout the photospheric regions with the applied vertical magnetic field mask. Figure~\ref{fig_vorticity} shows the \editors{results for different mask values}. We use the mask to exclude the contributions to the average value of the vorticity from the periphery, however, one should be careful with that because the number of contributing pixels is different at each time, and therefore the result can be highly affected by the emerging nature of the magnetic flux evolution. \editors{In particular, we believe that this is the cause for the significant initial drop of the average vertical vorticity seen in Figure~\ref{fig_vorticity} before $t=15$~min.} The snapshots of the vertical vorticity show that at the early stages, there are pixels with high vorticity values, but their total masked area (used for averaging) is still small, leading to the high values seen in the vorticity plot. As time is progressing and flux emerges, the area of masked pixels quickly increases, giving the obtained vorticity drop by $t=15$~min. Also, the analysis of the vertical magnetic field evolution shows that during the time interval before $t=15$~min, the magnetic polarities experience significant expansion in all directions \editors{without rotations}. This can explain very good correlations for both horizontal flow components during that interval. After $t=15$~min, the vorticity grows, or in other words, rotational motions become more pronounced, which affects the FLCT reconstructions. \editors{As seen from the plots for the 100~G and 200~G masks, the vorticity} saturates by $t=30$~min, that is why the reconstruction metrics stop decreasing, staying almost constant. After around 45~min \editors{(for the blue curve, and earlier for the yellow one)}, the vorticity decreases, so we can see an increase in the correlation coefficient both for $V_x$ and $V_y$ components. \editors{Concluding this Section, we however stress that this analysis of the dependence of the FLCT results on the vertical vorticity may not be considered as a strict explanation, because the vorticity plots presented in Figure~\ref{fig_vorticity} depend on the used strong-field mask.}}

\section{\pdfiss electric field inversions} \label{sec:pdfiss}
\begin{figure*}
\centering
\includegraphics[width=0.99\hsize, trim={0cm 0 0 0}, clip]{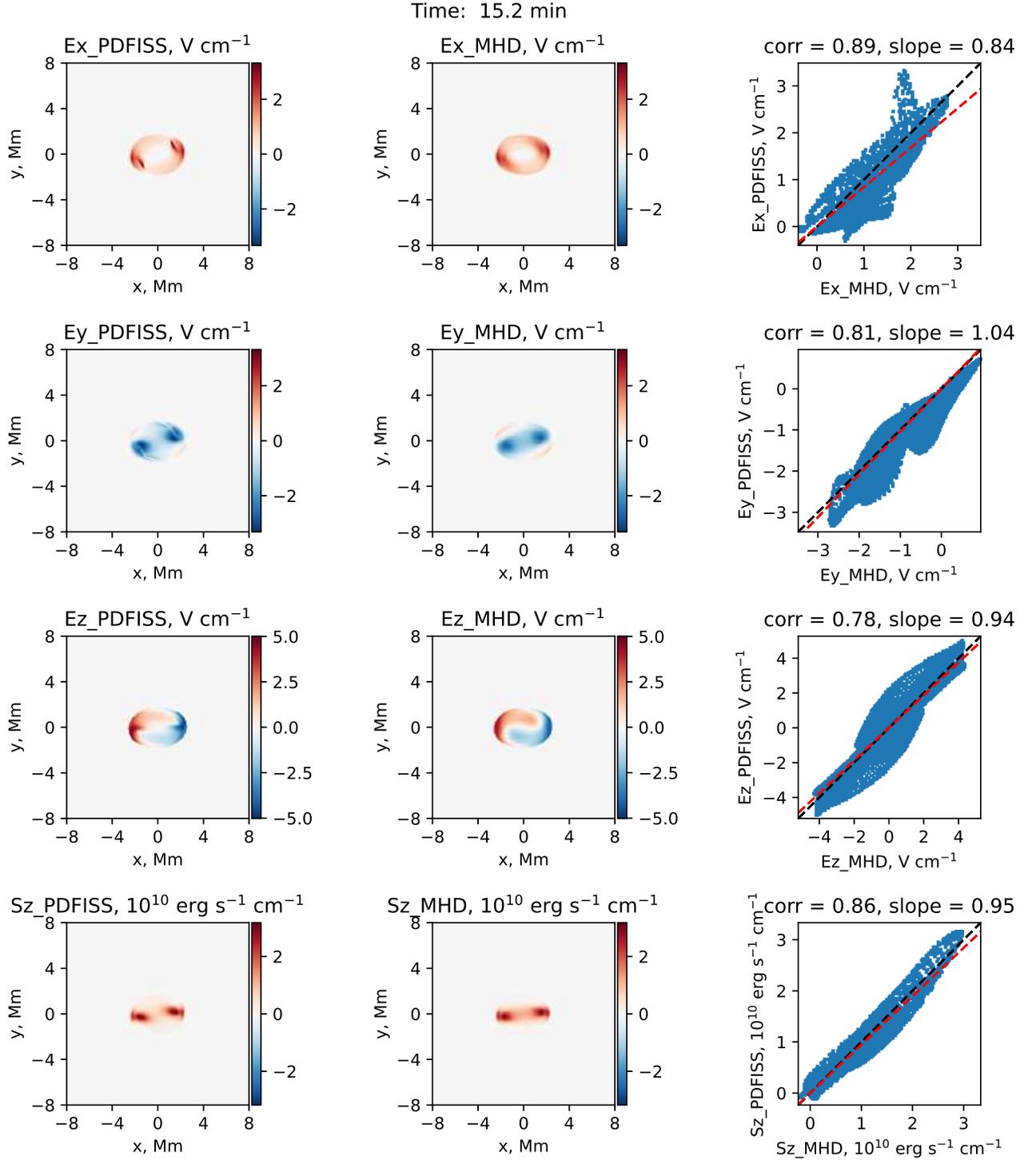}
\caption{Components of the electric field and vertical component of the Poynting flux obtained with \pdfiss method (left column) and the MFE simulation (middle column) at \andrei{$t=15.2$~min}, \benoit{\textit{i.e.} during the emergence phase of the simulation}. The right column shows the scatter plots with the linear fit (red dashed line), Spearman correlation coefficient and slope of the fit, and reference one-to-one line (black dashed).
\label{fig_efield_15}}
\end{figure*}

\begin{figure*}
\centering
\includegraphics[width=0.99\hsize, trim={0cm 0 0 0}, clip]{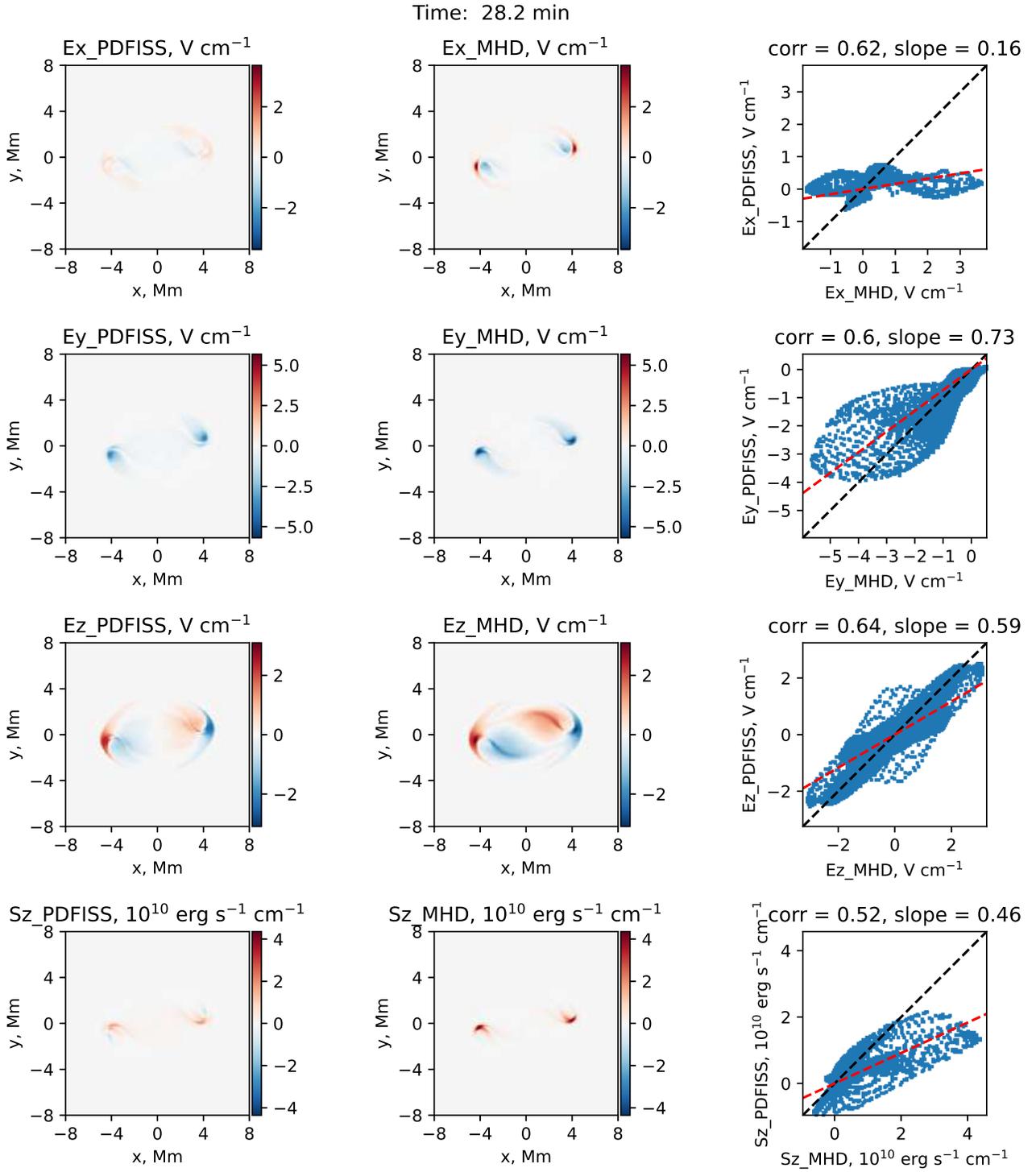}
\caption{The same as Figure~\ref{fig_efield_15} at \andrei{$t=28.2$~min}, \benoit{\textit{i.e.} during the early stages of the shearing phase of the simulation}.
\label{fig_efield_28}}
\end{figure*}

\begin{figure*}
\centering
\includegraphics[width=0.99\hsize, trim={0cm 0 0 0}, clip]{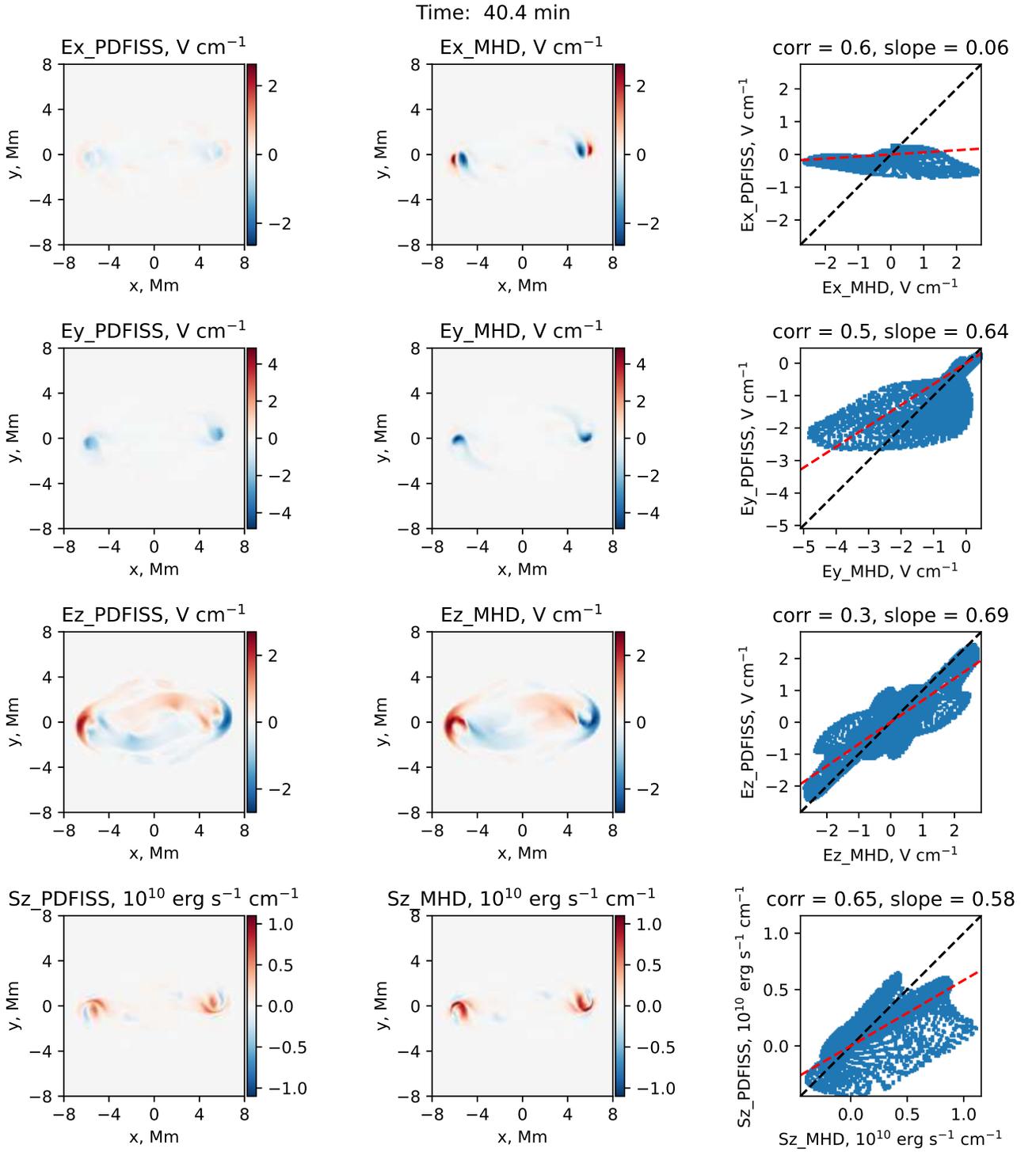}
\caption{The same as Figure~\ref{fig_efield_15} at \andrei{$t=40.4$~min}, \benoit{\textit{i.e.} during the later stages of the shearing phase of the simulation}.
\label{fig_efield_40}}
\end{figure*}

In this section, we use the \pdfiss method to reconstruct electric fields in the photosphere from the MFE simulated magnetograms and Dopplergrams, and FLCT estimates of the horizontal flows. Figures~\ref{fig_efield_15}, \ref{fig_efield_28}, and \ref{fig_efield_40} show the obtained electric fields and vertical component of the Poynting flux at three different times around 15, 28, and 40~min as well as the scatter plots comparing the reconstructions to actual MFE simulated values. We calculate the Spearman correlation coefficient and slope angle of the linear fit from the scatter plots. We find good agreement of the spatial structure of the inverted electric fields with the simulation data, although at the snapshots corresponding to the shearing phase ($t=28$ and 40~min snapshots), the $x$-component of the reconstructed electric field is highly underestimated.

\begin{figure*}
\centering
\includegraphics[width=0.6\hsize, trim={0cm 0 0 0}, clip]{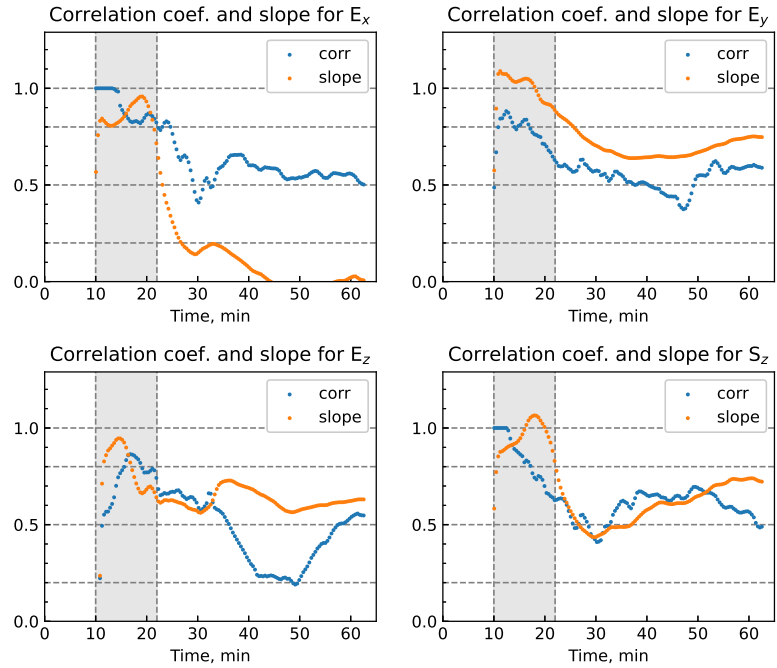}
\caption{Quality of the \pdfiss reconstruction for the entire duration of the simulation: Evolution of the Spearman correlation coefficient (blue) and slope parameter (yellow). The shaded area represents the emergence phase (10--22~min) of the simulated active region evolution. \benoit{The second dashed vertical line of each plot marks the beginning of the shearing phase ($t > 22$ min).} The horizontal dashed lines mark the limits for the strong, moderate, and weak correlation.  
\label{fig_pdfiss_corr_slope}}
\end{figure*}

Figure~\ref{fig_pdfiss_corr_slope} shows the evolution of the correlation coefficient and slope parameter for the three components of the electric field and the vertical component of the Poynting flux over the entire duration of the simulation. During the emergence phase (about 10--22~min, see Section~\ref{sec:mfe}), the \pdfiss results have strong-to-moderate correlation (0.5--1.0) with the MFE electric fields. The slope parameter is within 40\% difference with the ideal reconstruction unity value. During the shearing phase, the results have moderate-to-weak correlation (0.5--0.8), the slope is within 50\% difference for the $y$- and $z$-components, whereas the $x$-component of the electric field is significantly underestimated. We find low values for the correlation coefficient for the vertical component of the electric field at about 40--50~min. While the slope parameter is quite high (around 0.6--0.7), and the plotted maps show all the spatial structures seen in the actual MHD data (see e.g. Figure~\ref{fig_efield_40}), that correlation is unexpectedly low.

\begin{figure}
\centering
\includegraphics[width=0.89\hsize, trim={0cm 0 0 0}, clip]{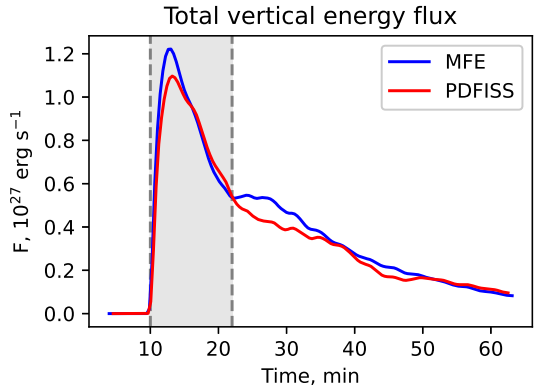}
\caption{Total vertical energy flux as computed by the MFE simulation (blue line) and reconstructed using \pdfiss (red line). The shaded area represents the emergence phase (10--22~min) of the simulated active region evolution. \benoit{The second dashed vertical line marks the beginning of the shearing phase ($t > 22$ min).}
\label{fig_pdfiss_total_energy_flux}}
\end{figure}

The reconstruction of the vertical Poynting flux is good during the emergence phase, with strong-to-moderate correlation and a slope within 20\% of the ideal reconstruction. During the shearing phase, the method still adequately reconstructs the vertical Poynting flux with moderate correlation except for a short time interval close to 30~min, and a slope within 60\% of the ideal unity value. We note that despite the significant underestimation and very low value of the slope for $E_x$, the vertical Poynting flux reconstruction quality is good. In addition, we calculate the total vertical energy flux, integrating spatially the PDFI\_SS-inverted vertical Poynting flux, as shown in Figure~\ref{fig_pdfiss_total_energy_flux}. The reconstructed total energy flux shows good agreement (within 20\%) with the MFE values. 

\begin{figure*}
\centering
\includegraphics[width=0.6\hsize, trim={0cm 0 0 0}, clip]{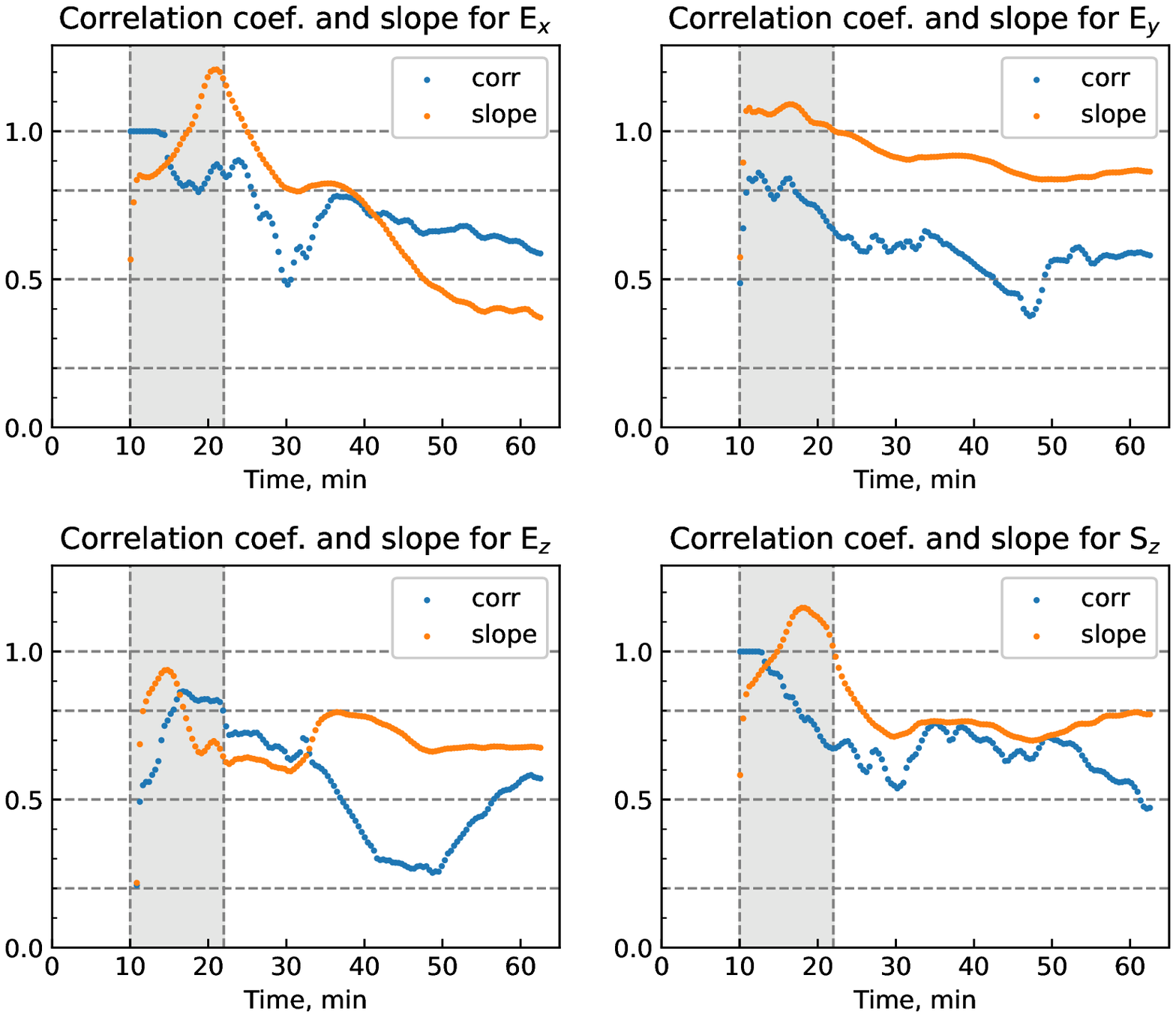}
\caption{Evolution of the Spearman correlation coefficient (blue) and slope parameter (yellow) of the \pdfiss inversions with actual MHD horizontal velocities instead of FLCT ones. The shaded area represents the emergence phase (10--22~min) of the simulated active region evolution. \benoit{The second dashed vertical line marks the beginning of the shearing phase ($t > 22$ min).} The horizontal dashed lines mark the limits for the strong, moderate, and weak correlation.
\label{fig_mfeVXVY_pdfiss_corr_slope}}
\end{figure*}

The poor value of the slope for the $x$-component of the electric field requires some additional attention. The discrepancy could result either from the structures in $V_y$ which are not detected in $B_z$ maps (see Section~\ref{sec:flct}) or from the more sophisticated nature of the active region evolution, including the shearing phase, which was not pronounced in the ANMHD test. To shed some light, we have run the \pdfiss software using actual MHD horizontal plasma velocities instead of the FLCT estimates. The obtained reconstruction (Figure~\ref{fig_mfeVXVY_pdfiss_corr_slope}) shows significantly better slope values for $E_x, E_y$, and $S_z$, but is still far from the close-to-ideal reconstruction (cf. ANMHD case, \citealp{Masha-2014ApJ...795...17K}). We therefore conclude that the poor FLCT reconstruction of the horizontal plasma flows due to the $V_y$ shear structures is not the only source of the discrepancy, although it does affect the electric field inversions. The shearing nature of the later stage of the simulated active region evolution comes into play. We speculate that during the shearing phase some discrepancy is due to the fact that the \pdfiss method misses the non-inductive (i.e. potential) contribution into the vertical component of the electric field. Another improvement would be to include additional observational information capturing the rotation of magnetic polarities in order to reconstruct the shear velocity structures properly. \andrei{Indeed, the simplified simulation of the flux emergence without taking into account the effect of convection results in a very smooth shape of polarity flux concentrations. By using other proxies in the FLCT technique \editors{or/and other velocity inference methods}, we could reconstruct better the shear velocity structures in the magnetic polarity regions, and therefore improve the quality of the \pdfiss reconstructions of the electric field (cf. Figure~\ref{fig_mfeVXVY_pdfiss_corr_slope}).}

\section{Conclusion} \label{sec:conclusion}

In this study, we performed a test to validate the \pdfiss electric field inversion method \citep{Fisher2020-ApJS..248....2F}, using the photospheric data extracted from an \yuhong{interior-to-corona MHD simulation of the emergence of a twisted flux tube}. Unlike the previous validation tests for the \pdfiss method that used just one snapshot of a rising flux tube in the solar interior (see, e.g. \citealp{Masha-2014ApJ...795...17K}), in this study we considered the more complex time sequence of evolution of the emerging flux region at the photosphere. The evolution of the simulated ephemeral active region consisted of two phases during which the vertical energy flux was dominated either by vertical plasma motions (the emergence phase) or by horizontal motions (the shearing phase). The initial rapid growth of the magnetic flux through the simulated photosphere corresponded to the emergence phase.

We used synthetic magnetograms and Dopplergrams at the photospheric level to calculate the inductive part of the photospheric electric field as well as its non-inductive (potential) parts due to the contributions from vertical and horizontal plasma flows, and ideal Ohm's law. The horizontal optical flows were estimated with the FLCT method \citep{FLCT2008ASPC..383..373F} tracking the evolution of the vertical component of the photospheric magnetic field.

Our main findings are the following:

\begin{itemize}

\item During the emergence phase of the simulation, the \pdfiss reconstruction of the photospheric electric field is very good -- the correlation between the reconstructed and MFE simulation fields varies from strong to moderate. During the shearing phase, the accuracy of the \pdfiss method decreases -- the correlation between the reconstructed and simulation fields is moderate-to-weak, with the fields, especially the $E_x$ component, being underestimated. 

\item The \pdfiss method reconstructs the vertical component of the Poynting flux well. The correlation between the reconstructed and MFE simulation energy fluxes varies mostly from moderate to strong. The spatially integrated total energy flux is within 20\% of the simulation flux throughout both the emergence and shearing phases of the active region evolution. We note that reliable inversions of the energy flux is of special importance for data-driven MHD simulations of solar eruptions.

\item Difficulties in reconstructing the $y$-component of the horizontal velocity using the FLCT method, lead to a decrease in accuracy for the electric field reconstruction during the shearing phase of the AR evolution. We conclude that \editors{these difficulties are} caused by the lack of spatial structure in symmetric rotating sunspots as seen in $B_z$ magnetograms used as an input for the FLCT, \editors{while the $x$-component of the horizontal velocity is reconstructed better due to strong contributions from the translational motions of the magnetic polarities at the opposite directions}. We speculate that the \editors{electric field} reconstruction accuracy is also affected by the lack of the non-inductive contribution to the vertical component of the electric field. 

\item To test whether the \pdfiss method can adequately describe the simulated MHD electric fields, we have performed the reconstructions using the MFE simulation horizontal plasma velocities (instead of the FLCT velocities). In this case, the \pdfiss method yields more accurate electric fields, especially for $E_x$.  However, the reconstructed electric fields still differ from MFE data. We believe that the complex nature of the evolution of the simulated active region that includes the shearing evolution phase, as compared to the ANMHD case, does not allow the \pdfiss method to provide the highest quality reconstruction of the photospheric electric field. 

\end{itemize}

The flux emergence simulation used for this validation is still very idealised in that it does not include the effect of convection so that the emerging flux pattern consists of very smooth polarity flux concentrations. This makes the detection of the shear and rotational motions more difficult than reality and thus may be an underestimate of the sensitivity of the \pdfiss method. \andrei{We also note that the reliable detection of the velocities of rotating structures is a fundamental problem affecting all  inversion and tracking methods that depend on observed time variability. For instance, a rotating sunspot that has perfect symmetry about the axis will show no evolution and hence zero electric and velocity field even if the actual field is substantial. In this case, however, real observations offer an advantage. Many examples of rotating sunspots show structures in white light and sometimes in magnetic fields, which clearly rotate, and correlation tracking methods could capture these motions \citep{Masha-2009ApJ...704.1146K}.}

\andrei{To summarise, the accuracy of the \pdfiss reconstructions of the electric fields decreases for the shearing evolution of the active region, however, the spatial structure of electric fields is captured correctly, correlations are still decent, and the energy flux is reconstructed very well. As mentioned previously, the information on the photospheric electric fields is of special importance for cutting-edge data-driven MHD simulations. This field of physics of the solar atmosphere still contains a lot of uncertainties (see, for instance, the recent comparison study by \citet{Toriumi_2020} ) and needs yet to formulate its general methodology. In this light, the information delivered by the \pdfiss method is of crucial importance and \editors{undoubtedly} will be used in data-driven MHD simulations of solar eruptive processes.}

\acknowledgments
We acknowledge support of NASA ECIP NNH18ZDA001N and NASA LWS NNH17ZDA001N (A.N.A. and M.D.K.), NASA LWS 80NSSC19K0070 (Y.F.), DKIST Ambassador Program (A.N.A.), and Program of Basic Research No. II.16.1.6 (A.N.A.). 

\software{\pdfiss \citep{Fisher2020-ApJS..248....2F}, MFE \citep{Fan:2017}, FLCT \citep{FLCT2008ASPC..383..373F}}

\bibliographystyle{aasjournal}
\bibliography{biblio-pdfiss-paper}

\end{document}